\documentclass[aps,prd,twocolumn,showpacs]{revtex4}
\begin{document}
\title{Unification of standard and exotic matter
       through a $\bf Z_2$ symmetry}
\author{Ernesto A. Matute}
\email{ematute@lauca.usach.cl (E.A. Matute).}
\affiliation{Departamento de F\'{\i}sica, Universidad de Santiago
de Chile, Casilla 307, Santiago 2, Chile}
\begin{abstract}
We consider a scenario in which the discrete weak symmetry between
quarks and leptons is extended to the weak force by introducing
exotic partners. We conjecture that there exists a hidden discrete
symmetry $\tilde{\mbox{P}}$, defining a ${\bf Z_2}$ group, between
standard and exotic quarks and leptons. The unified model
$\mbox{SU(3)}_{q} \times \mbox{SU(3)}_{\tilde{q}} \times
\mbox{SU(2)}_{q \tilde{\ell}} \times \mbox{SU(2)}_{\tilde{q} \ell}
\times \mbox{U(1)}_{Y} \times \tilde{\mbox{P}}$ is discussed,
where the unifying discrete symmetry extends over particles and
forces. It is shown that the lighter neutral and charged weak
bosons generated upon spontaneous symmetry breaking have the same
properties as those of the standard model.
Cabbibo-Kobayashi-Maskawa unitarity is used to set a mass of order
2.8 TeV for the nonstandard weak bosons, which do not exhibit
quark-lepton universality.  A grand unified theory of
$(\mbox{GUT})_{q \tilde{\ell}} \times (\mbox{GUT})_{\tilde{q}
\ell} \times \tilde{\mbox{P}}$ type, with decay of exotic matter
into standard matter and no decay of the ordinary nucleon, is put
forward.
\end{abstract}
\pacs{12.60.Cn; 11.30.Hv; 12.10.Dm; 14.70.Pw}
\maketitle

The unification of quark and lepton properties under strong and
electroweak interactions through a continuous or discrete symmetry
has been one of the targets of research efforts over the last few
decades~\cite{Mohapatra}. Mainly the idea has been to assign
similar degrees of freedom to quarks and leptons, extending in a
sense the color gauge group of quarks to include leptons.
Nonstandard or exotic gauge bosons and fermions are commonly added
to the spectrum of the standard model (SM).  The purpose of this
paper is to implement these developments with the notion that
there is a hidden ${\bf Z_2}$ symmetry in nature unifying the
properties of standard and exotic fermions. And that these
interact with separate but symmetric gauge bosons, so duplicating
the gauge group $\mbox{SU(3)}_{q} \times \mbox{SU(2)}_{L} \times
\mbox{U(1)}_{Y}$ of the SM and its unified extensions.  In the
scenario of a duplicated grand unified theory (GUT) with such a
discrete symmetry, the decay of exotic matter into standard matter
and the stability of the ordinary nucleon are expected. This makes
an interesting point as nucleon decay is one of the pillars of
GUTs and no ordinary proton decay has yet been observed.

In the strong interaction sector of the SM there are substantial
differences between quarks and leptons as they come in triplets
and singlets of the color gauge group $\mbox{SU(3)}_{q}$,
respectively. However, in the electroweak sector their properties
are so similar that it is conceivable an interaction gauge
symmetry completely or at least partially duplicated in the form
$\mbox{G}_{q} \times \mbox{G}_{\ell}$ but unified through a
discrete symmetry; the subscripts $q$ and $\ell$ are to emphasize
that each $G$ only acts on quarks or leptons. A trouble with this
picture is that the $\mbox{U(1)}_{Y}$ charges of quarks and
leptons are different.  Nonetheless, the symmetric duplication of
the electroweak gauge group produces triangle anomalies which can
be cancelled by introducing exotic quarks and leptons, related to
the standard ones through the same discrete symmetry.  This
pattern then allows to expand the standard electroweak gauge group
to the unified form $\mbox{G}_{q \tilde{\ell}} \times
\mbox{G}_{\tilde{q} \ell} \times \tilde{\mbox{P}}$, where
$\tilde{q}$ and $\tilde{\ell}$ denote exotic quarks and exotic
leptons, respectively, and $\tilde{\mbox{P}}$, hereafter referred
to as exotic symmetry, denotes the discrete symmetry which
transforms standard quarks and leptons to exotic ones and vice
versa: $q \leftrightarrow \tilde{q}$, $\ell \leftrightarrow
\tilde{\ell}$. This unifying symmetry defines a ${\bf Z_2}$ group
as $\tilde{\mbox{P}}^2=1$. Moreover, this discrete exotic symmetry
is extendible to strong interactions.

Our aim in this paper is to discuss the unified gauge group
$\mbox{SU(3)}_{q} \times \mbox{SU(3)}_{\tilde{q}} \times
\mbox{SU(2)}_{q \tilde{\ell}} \times \mbox{SU(2)}_{\tilde{q} \ell}
\times \mbox{U(1)}_{Y} \times \tilde{\mbox{P}}$, assuming that
each set of fermions is separately conserved.  It is a unified
model in the sense that the duplicated gauge group factors have
the same gauge coupling constant, and the standard and exotic
quarks and leptons have similar quantum numbers.  Symmetric
standard and exotic quarks interact with symmetric gluons and weak
bosons. The two $\mbox{SU(3)}$ color groups are unbroken and
exotic gluons confine exotic quarks into heavy exotic hadrons. Any
explanation for the mass hierarchy of standard quarks should be
extended to include exotic quarks, since standard and exotic
gluons confine at the same scale; in a sense, three new heavy
generations of quarks and leptons are added to the standard
spectrum.  We may say that the exotic symmetry $\tilde{\mbox{P}}$
has the incentive of expanding the electroweak quark-lepton
symmetry to the strong interaction sector.  We note here that the
spontaneous breakdown of the discrete symmetry $\tilde{\mbox{P}}$
leads to well-known cosmological problems associated with
topological defects, namely, domain walls; they can be solved,
however, via interactions of domain walls with primordial black
holes~\cite{SFS}.  The model provides a framework for
understanding possible deviations from the SM predictions. We use
the reported quark-lepton universality violation of weak
interactions at low energies to fix the scale of the new weak
physics.

Specifically, the transformation properties of standard quarks and
leptons are
\begin{eqnarray}
& q^{i}_{nL} \sim (3, 1, 2, 1, \frac{1}{3}), \;\;\;\;
\ell_{nL} \sim (1, 1, 1, 2, -1) , & \nonumber \\
& u^{i}_{nR} \sim (3, 1, 1, 1, \frac{4}{3}) , \;\;\;\;
\nu_{nR} \sim (1, 1, 1, 1, 0) , & \nonumber \\
& d^{i}_{nR} \sim (3, 1, 1, 1, - \frac{2}{3}) , \;\;\;\; e_{nR}
\sim (1, 1, 1, 1, -2) , & \label{fermions}
\end{eqnarray}
where $n$ denotes the three generations, $q_{nL}=(u_{nL},d_{nL})$
and $\ell_{nL}=(\nu_{nL},e_{nL})$ are the $\mbox{SU(2)}$ doublets,
$i$ denotes the color degree of freedom, and the numbers in
brackets describe the transformation properties under the unified
gauge group.  The $\nu_{R}$'s are introduced because of the
experimental signatures for neutrino masses~\cite{PDG}.

The quark and lepton assignments in Eq.~(\ref{fermions}), however,
introduce axial anomalies of the $\mbox{U(1)}_{Y}[\mbox{SU(2)}]^2$
type. They are associated with the nonvanishing trace of $Y$ for
quarks and for leptons due to the separation of the weak gauge
group. To achieve their cancellation we include exotic fermions as
done in the so-called ununified models~\cite{Georgi}. We consider
the exotic quarks and leptons
\begin{eqnarray}
& \tilde{q}^{i}_{nL} \sim (1, 3, 1, 2, \frac{1}{3}) , \;\;\;\;
\tilde{\ell}_{nL} \sim (1, 1, 2, 1, -1) , & \nonumber \\
& \tilde{u}^{i}_{nR} \sim (1, 3, 1, 1, \frac{4}{3}) , \;\;\;\;
\tilde{\nu}_{nR} \sim (1, 1, 1, 1, 0) , & \nonumber \\
& \tilde{d}^{i}_{nR} \sim (1, 3, 1, 1, - \frac{2}{3}) , \;\;\;\;
\tilde{e}_{nR} \sim (1, 1, 1, 1, -2) . &
\label{exotics}
\end{eqnarray}
Under the exotic symmetry $\tilde{\mbox{P}}$, ordinary and exotic
fermions are transformed as follows:
\begin{eqnarray}
& q^{i}_{nL} \leftrightarrow \tilde{q}^{i}_{nL} , \;\;\;
u^{i}_{nR} \leftrightarrow \tilde{u}^{i}_{nR} , \;\;\;
d^{i}_{nR} \leftrightarrow \tilde{d}^{i}_{nR} , & \nonumber \\
& \tilde{\ell}_{nL} \leftrightarrow \ell_{nL} , \;\;\;
\tilde{\nu}_{nR} \leftrightarrow \nu_{nR} , \;\;\;
\tilde{e}_{nR} \leftrightarrow e_{nR} , &
\label{exoticsym}
\end{eqnarray}
and $W^{a}_{q} \leftrightarrow W^{a}_{\tilde{q}}$, $G^{b}_{q}
\leftrightarrow G^{b}_{\tilde{q}}$, where $W^{a}_{q}$,
$W^{a}_{\tilde{q}}$ are the weak gauge bosons of $\mbox{SU(2)}_{q
\tilde{\ell}}$, $\mbox{SU(2)}_{\tilde{q} \ell}$, and $G^{b}_{q}$,
$G^{b}_{\tilde{q}}$ are the gluons of $\mbox{SU(3)}_{q}$,
$\mbox{SU(3)}_{\tilde{q}}$, respectively. Besides, the
corresponding bare gauge couplings are required to be equal. Thus
the whole Lagrangian of electroweak and strong interactions, with
the Higgs fields defined below, becomes invariant under
$\tilde{\mbox{P}}$ symmetry. This leads to a great simplification
of many results which cannot be achieved for ununified models due
to current structures; phenomenological implications are therefore
different.

In this work we assume that the weak gauge symmetry and exotic
symmetry $\tilde{\mbox{P}}$ are spontaneously broken at one stage,
which happens with our choice of Higgs multiplets. The minimal set
of colorless scalar multiplets that breaks the electroweak gauge
symmetry $\mbox{SU(2)}_{q \tilde{\ell}} \times
\mbox{SU(2)}_{\tilde{q} \ell} \times \mbox{U(1)}_{Y} \times
\tilde{\mbox{P}}$ down to $\mbox{U(1)}_{em}$ and gives mass to all
fermions of the model includes
\begin{eqnarray}
& & \phi_{q} \sim (2, 1, +1) ,  \; \; \;\; \phi_{\tilde{q}} \sim
(1, 2, +1) , \nonumber \\ & & \Sigma = \sigma + i \mbox{\boldmath
$\tau$} \cdot \mbox{\boldmath $\pi$} \sim (2, 2, 0) ,
\label{Higgs}
\end{eqnarray}
with vacuum expectation values (VEVs)
\begin{eqnarray}
& & \langle \phi_{q} \rangle = \left(
\begin{array}{c}
0 \\  \\  v / \sqrt{2} \end{array} \right) , \;\;\;\; \langle
\phi_{\tilde{q}} \rangle = \left(
\begin{array}{c} 0 \\  \\  v / \sqrt{2} \end{array} \right) ,
\nonumber \\
& & \langle \Sigma \rangle = \left(
\begin{array}{cc} u & 0 \\  \\ 0 & u \end{array} \right) .
\label{VEVS}
\end{eqnarray}
The Pauli matrices $\mbox{\boldmath $\tau$}$ in Eq.~(\ref{Higgs})
belong to the standard gauge group $\mbox{SU(2)}_{L}$ and they are
there to indicate that after spontaneous symmetry breaking
$\Sigma$ splits into a singlet and a triplet. Each row and each
column of $\Sigma$ is a doublet under $\mbox{SU(2)}_{q
\tilde{\ell}}$ and $\mbox{SU(2)}_{\tilde{q} \ell}$, respectively.
Under $\tilde{\mbox{P}}$ symmetry $\phi_{q} \leftrightarrow
\phi_{\tilde{q}}$ and $\Sigma \leftrightarrow \Sigma$.  The
equality of the bare VEVs of $\phi_{q}$ and $\phi_{\tilde{q}}$ is
broken by radiative corrections due to the hierarchy of Yukawa
couplings responsible for fermion masses, so that the renormalized
quantities are actually equal up to finite radiative corrections.
We remark that this type of Higgs multiplets is also present in
ununified models, but without any restriction on the VEVs of Higgs
doublets.

The gauge symmetry breaking follows the pattern
\begin{eqnarray}
& \mbox{SU(2)}_{q \tilde{\ell}} \times \mbox{SU(2)}_{\tilde{q}
\ell} \times \mbox{U(1)}_{Y} \times \tilde{\mbox{P}}
& \nonumber \\ & \downarrow & \nonumber \\
& \mbox{SU(2)}_{L} \times \mbox{U(1)}_{Y} & \nonumber \\
& \downarrow & \nonumber \\ & \mbox{U(1)}_{em} . &
\end{eqnarray}
The unbroken electric charge generator is given by
\begin{equation}
Q = I_{3L} + \frac{1}{2} Y = I_{3q} + I_{3\tilde{q}} + \frac{1}{2}
Y ,
\end{equation}
and the electroweak couplings of the model, in terms of the
standard electromagnetic coupling constant and Weinberg angle,
take the form
\begin{equation}
g = g_{q} = g_{\tilde{q}} = \frac{\sqrt{2} e}{\sin \theta_{W}} ,
\;\;\; g' = \frac{e}{\cos \theta_{W}} . \label{gs}
\end{equation}

The two physical sets of left-handed weak bosons and the photon
get the following linear combinations of the electroweak gauge
bosons $W_{q}$, $W_{\tilde{q}}$ and $A$:
\begin{eqnarray}
& & W^{\pm}_{1} = \frac{1}{\sqrt{2}} (W^{\pm}_{q} +
W^{\pm}_{\tilde{q}}) , \nonumber \\ & & Z_{1} = \frac{1}{\sqrt{2}}
\cos \theta_{W} (W^{3}_{q} + W^{3}_{\tilde{q}}) - \sin \theta_{W}
A ,
\nonumber \\
& & A_{em} = \frac{1}{\sqrt{2}} \sin \theta_{W} (W^{3}_{q} +
W^{3}_{\tilde{q}}) + \cos \theta_{W} A ,
\nonumber \\
& & W^{\pm}_{2} = \frac{1}{\sqrt{2}} (W^{\pm}_{q} -
W^{\pm}_{\tilde{q}}) ,
\nonumber \\
& & Z_{2} = \frac{1}{\sqrt{2}} (W^{3}_{q} - W^{3}_{\tilde{q}}) .
\label{gaugebosons}
\end{eqnarray}
They pick up the mass
\begin{eqnarray}
& & m_{W_{1}}^{2} = \frac{1}{4} g^{2} v^{2} ,  \;\;\; m_{Z_{1}}^{2}
= \frac{1}{4} (g^{2} + 2 g'^{2}) v^{2} \; = \;
\frac{m_{W_{1}}^{2}}{\cos^{2} \theta_{W}} , \nonumber \\ & &
m_{A_{em}} = 0 , \;\;\;  m_{W_{2}}^{2} = \frac{1}{4} g^{2} (v^{2}
+ 2 u^{2}) , \nonumber \\ & & m_{Z_{2}} = m_{W_{2}} .
\label{mass}
\end{eqnarray}
The gauge fields ${\bf W}_{1} = ({\bf W}_{q}+{\bf W}_{\tilde{q}})
/ \sqrt{2}$ are identified with the SM weak bosons, while the
orthogonal combination ${\bf W}_{2} = ({\bf W}_{q} - {\bf
W}_{\tilde{q}}) / \sqrt{2}$ represents the nonstandard ones. We
note that the standard mass relation $m_{W_{1}} = m_{Z_{1}} \cos
\theta_{W}$ is preserved. Also significant, the heavier
nonstandard neutral and charged weak bosons acquire the same mass.
All of these only occurs within the context of the unified model
of ordinary and exotic quarks and leptons in which the exotic
symmetry $\tilde{\mbox{P}}$ is essential, making the difference
from other extensions of the SM.

The charged-current interactions involving the $W_{1}^{\pm}$ and
$W_{2}^{\pm}$ left-handed bosons can be written as
\begin{equation}
{\cal L}_{CC} = \frac{g}{2} (W_{1\mu}^{+} J_{1L}^{-\mu} +
W_{2\mu}^{+} J_{2L}^{-\mu} + \mbox{h.c.}) ,
\label{CC}
\end{equation}
with weak currents defined, at the ordinary and exotic quark and
lepton level, by
\begin{eqnarray}
J_{1L}^{\pm} &=& J_{qL}^{\pm} + J_{\ell L}^{\pm}
+ J_{\tilde{\ell}L}^{\pm} + J_{\tilde{q}L}^{\pm}  ,
\nonumber \\
J_{2L}^{\pm} &=& J_{qL}^{\pm} - J_{\ell L}^{\pm}
+ J_{\tilde{\ell}L}^{\pm} - J_{\tilde{q}L}^{\pm}  .
\label{charcurr}
\end{eqnarray}
As for the neutral-current interactions involving the $Z_{1}$ and
$Z_{2}$ left-handed bosons, we have
\begin{eqnarray}
{\cal L}_{NC} &=& \frac{g}{\sqrt{2} \cos \theta_{W}} Z_{1\mu}
(J_{1L}^{o\mu} - \sin^{2} \theta_{W} J_{em}^{\mu}) \nonumber \\ &
& + \frac{g}{\sqrt{2}} Z_{2\mu} J_{2L}^{o\mu} ,
\label{NC}
\end{eqnarray}
where now
\begin{eqnarray}
J_{1L}^{o} &=& J_{qL}^{3} + J_{\ell L}^{3} + J_{\tilde{\ell}L}^{3}
+ J_{\tilde{q}L}^{3} ,  \nonumber \\ J_{2L}^{o} &=& J_{qL}^{3} -
J_{\ell L}^{3} + J_{\tilde{\ell}L}^{3}
- J_{\tilde{q}L}^{3} ,  \nonumber \\
J_{em} &=& J_{qL}^{3} + J_{\ell L}^{3} + J_{\tilde{\ell}L}^{3} +
J_{\tilde{q}L}^{3}  \nonumber \\ & & + J_{qY} + J_{\ell Y} +
J_{\tilde{\ell} Y} + J_{\tilde{q} Y} .
\label{NCs}
\end{eqnarray}

We see, from Eqs.~(\ref{gs})--(\ref{NCs}), that $W_{1}^{\pm}$ and
$Z_{1}$ have the SM mass and coupling relationships with a
universality that comprises exotic fermions, and no mixing with
the nonstandard $W_{2}^{\pm}$ and $Z_{2}$ exists, which exhibit no
quark-lepton universality and also the breaking of exotic
symmetry. At the mass scale of the SM, this corresponds to adding
three new generations of quarks and leptons.

We next write the most general renormalizable, gauge invariant and
$\tilde{\mbox{P}}$ symmetric Lagrangian for the Yukawa couplings
of Higgs bosons with fermions, consistent with separate ordinary
and exotic baryon number conservation, as well as separate
ordinary and exotic lepton number conservation, which are
associated with U(1) global vector symmetries:
\begin{eqnarray}
{\cal L}_{YUK} &=& G_{nm}^{d} \left( \overline{q}_{nL} \phi_{q}
d_{mR} + \overline{\tilde{q}}_{nL} \phi_{\tilde{q}} \tilde{d}_{mR}
\right)
\nonumber \\
& & + G_{nm}^{u} \left( \overline{q}_{nL} \phi^{c}_{q} u_{mR} +
\overline{\tilde{q}}_{nL} \phi^{c}_{\tilde{q}} \tilde{u}_{mR}
\right)
\nonumber \\
& & + G_{nm}^{e} \left( \overline{\ell}_{nL} \phi_{\tilde{q}}
e_{mR} + \overline{\tilde{\ell}}_{nL} \phi_{q} \tilde{e}_{mR}
\right)
\nonumber \\
& & + G_{nm}^{\nu} \left( \overline{\ell}_{nL}
\phi^{c}_{\tilde{q}} \nu_{mR} + \overline{\tilde{\ell}}_{nL}
\phi^{c}_{q} \tilde{\nu}_{mR} \right) \nonumber \\ & & +
\mbox{h.c.} \label{Yukawa}
\end{eqnarray}
Because of gauge invariance, only the ordinary or exotic lepton
number conservation is related to an apparently \emph{ad hoc}
global symmetry. The left-handed bidoublet $\Sigma$ then has no
Yukawa couplings to fermions. Substitution of VEVs leads to mass
matrices that can be made real and diagonal by suitable unitary
transformations of the left- and right-handed components of the
pertinent fermion fields. In this mass eigenstate basis, $Z_{1}$
and $Z_{2}$ have flavor-conserving couplings.

The equality between Yukawa couplings of ordinary and exotic
particles yields, however, equal mass values at tree level. These
mass relationships can easily be avoided if two new Higgs doublets
$\phi'_{q} \sim (2, 1, -1)$ and $\phi'_{\tilde{q}} \sim (1, 2,
-1)$ are added, which under $\tilde{\mbox{P}}$ symmetry transform
as $\phi'_{q} \leftrightarrow \phi^{c}_{\tilde{q}}$ and
$\phi'_{\tilde{q}} \leftrightarrow \phi^{c}_{q}$.  It can be seen
that in this case the new Yukawa Lagrangian leads to no standard
and exotic mass relations if the VEVs of the new Higgs doublets
are equal to each other but different from those in
Eq.~(\ref{VEVS}). This complication in the Higgs sector has no
effect on the properties of the gauge bosons discussed in this
paper, except that their masses now receive contributions from the
new doublets (see Eq.~(\ref{mass})).  As mentioned above, all of
this happens because of the $\tilde{\mbox{P}}$ symmetry for the
full Lagrangian.  Clearly, this scenario, rendering the model
rather uneconomic but richer phenomenologically, favors the SM
with two Higgs doublets.

Unfortunately, the systematics of the masses of the known quarks
and leptons are unexplained, so that there is no reliable
extrapolation to masses of exotic generations.  However, strict
constraints on the masses of the new fermions are obtained from
bounds placed by cosmology and precision electroweak experiments.

On the one side, cosmological evidences on the ordinary baryon
asymmetry and absence at present of exotic baryons made of exotic
quarks require that exotic matter be unstable, decaying rapidly
enough into observable ordinary matter~\cite{PDG}. As argued
below, this is a strong motivation to embed the duplicated SM into
a duplicated GUT such that exotic quarks and ordinary leptons
transform under one GUT gauge group, and exotic leptons and
ordinary quarks do under the other. Besides of allowing the decay
of exotic matter, such a duplicated GUT would indicate the
ordinary proton stability. In this scenario, the cosmological
ordinary baryon (lepton) asymmetry requires the exotic lepton
(baryon) asymmetry. Thus, the problem of ordinary baryogenesis
(leptogenesis) is also that of exotic leptogenesis
(baryogenesis)~\cite{Buchmuller}.

On the other side, there are constraints on the masses and
mass-splittings of exotic fermions from the standard $Z$ and $W$
gauge boson observables. If any of the exotic fermions is lighter
than about half of the mass of the SM-like $Z$ boson of the model,
this will affect its decay width very strongly. For instance, if
the $Z$ can decay into exotic light neutrinos, its width  will be
significantly increased, in contradiction with experiment.  In
fact, the direct measurement of the invisible $Z$ width gives the
bound $N_{\nu}=2.92 \pm 0.07$ on the number of light neutrino
types~\cite{PDG}. On the contrary, if the exotic fermions are
heavy, their couplings to the lighter $Z$ and $W$ bosons will
affect the values of the precision electroweak   \linebreak
observables $S$, $T$ and $U$, or equivalent oblique parameters on
the gauge self-energies at the loop level~\cite{Peskin}.  These
parameters are defined to constrain heavy new physics, vanishing
for the SM. In the case of new generations of fermions, these
values restrain the masses and mass-splittings of such fermions.
For completeness, we present in the following the contributions to
these oblique parameters from each nondegenerate doublet
$(f_{1},f_{2})$ of extra SM-like heavy quarks or leptons with
masses $(m_{1},m_{2})$, in the limit $m_{Z}, \Delta m =
|m_{1}-m_{2}| \ll m_{1,2}$ \cite{Peskin}:
\begin{eqnarray}
S & \simeq & \frac{N_{c}}{6 \pi} , \nonumber \\
T & \simeq & \frac{N_{c}}{12 \pi \sin^{2}\theta_{W}
\cos^{2}\theta_{W}} \; \frac{\Delta m^{2}}{m_{Z}^{2}} ,
\nonumber \\
U & \simeq & \frac{2 N_{c}}{15 \pi} \; \frac{\Delta
m^{2}}{m_{1}^{2}} ,
\end{eqnarray}
where $N_{c}=3 \; (1)$ denotes the color number of extra quarks
(leptons). We note that if $m_{1}$ and $m_{2}$ are nondegenerate
and not too large, additional negative corrections to $S$ may
arise; this parameter in a sense measures the size of the new
fermion sector, while $T$ and $U$ do on its weak-isospin breaking.
The experimental values for these observables are~\cite{PDG}
\begin{eqnarray}
& & S = - 0.13 \pm 0.10 \; (- 0.08) , \nonumber \\
& & T = - 0.17 \pm 0.12 \; (+ 0.09) , \nonumber \\
& & U = 0.22 \pm 0.13 \; (+ 0.01) ,
\end{eqnarray}
where the central values assume $m_{H}$=117 GeV for the SM Higgs
boson, and the values shown in the parentheses correspond to the
changes for $m_{H}$=300 GeV; the $S$ and $T$ parameters are highly
correlated. It has been shown that three generations of relatively
heavy extra quarks and leptons cannot fit the data with the SM
Higgs doublet, but they do it with a two-Higgs-doublet extension,
although the parameter space becomes very restrictive with fermion
masses significantly below 1 TeV~\cite{He}; the nonstandard gauge
bosons would have an effect on the oblique parameters only at the
two-loop level through their couplings to fermions, because as
exhibited in Eq.~(\ref{gaugebosons}) they do not mix with the
standard ones~\cite{Wells}. This motivates even more, within the
context of our model, the duplication of the SM forces with a
two-Higgs-doublet extension included.

Hence, all of these can be met if the exotic fermions are massive
and nondegenerate enough to decay into ordinary fermions in a
duplicated GUT scenario, with a generic lower mass bound about the
standard $m_{Z}/2$. By contrast, if the masses are significantly
greater than the generic bound $\langle H \rangle \simeq$ 174 GeV,
the Higgs VEV causing the SM-like electroweak symmetry breaking,
then the weak interaction among heavy fermions becomes strong and
perturbation theory breaks down, so eluding any mass bound
depending on the perturbative assumption. Collider searches for
new long-lived up and down quarks give by now the lower mass
bounds of 220 GeV and 190 GeV, respectively, while the current
direct lower bounds on extra stable neutral and charged leptons
are about 45.0 GeV and 102.6 GeV, respectively \cite{PDG}.
Moreover, the ordinary and exotic fermions do couple to the
degenerate heavy nonstandard $Z$ and $W$ gauge bosons, so that the
new physics cannot be fully parametrized in the usual $S$, $T$ and
$U$ framework.  Actually, we will use the mass ratio between the
light and heavy $W$ (see Eq.~(\ref{eta}) below).  And to restrain
this new parameter, we will consider in this paper charged-current
and neutral-current data at the standard matter level.

The relevant charged-current effective Lagrangian for the
low-energy ordinary quark-lepton physics becomes
\begin{equation}
{\cal L}^{\mbox{eff}}_{\mbox{CC}} = \frac{4 G_{F}}{\sqrt{2}
(1+\eta)} (J^{+}_{1L\mu} J^{-\mu}_{1L} + \eta J^{+}_{2L\mu}
J^{-\mu}_{2L}) , \label{eff}
\end{equation}
where $J^{\pm}_{1,2 L}$ are the currents in Eq.~(\ref{charcurr})
limited to standard quarks and leptons, $\eta$ is a parameter
defined in the minimal model by
\begin{equation}
\eta = \frac{m^{2}_{W_{1}}}{m^{2}_{W_{2}}} =
\left( 1 + \frac{2 u^{2}}{v^{2}} \right)^{-1} ,
\label{eta}
\end{equation}
and $G_{F}$ is the Fermi constant from the muon decay determined
(at tree level) by
\begin{equation}
\frac{4 G_{F}}{\sqrt{2}} = \frac{g^{2}}{4 m^{2}_{W_{1}}} \;
(1 + \eta) .
\label{Fermi}
\end{equation}
According to Eq.~(\ref{eff}), there is no universality in the
strength of leptonic and hadronic weak interactions, which depends
on the magnitude of the $\eta$-parameter.  The corresponding
effective neutral-current weak interactions are given by an
expression similar to Eq.~(\ref{eff}), which is derived from
Eq.~(\ref{NC}) restricted to conventional quarks and leptons.

Therefore, within our scenario the electroweak predictions of the
SM also get corrections parameterized by the quantity $\eta$
defined in Eq.~(\ref{eta}).  To exhibit the degree of violation of
the universality of weak interactions, we consider the constraint
dictated by low-energy charged-current data.  At the low-energy
level of ordinary quarks and leptons, charged-current interactions
are governed by the effective Lagrangian of Eq.~(\ref{eff}), which
can be rewritten as follows:
\begin{eqnarray}
{\cal L}^{\mbox{eff}}_{\mbox{CC}} &=& \frac{4 G_{F}}{\sqrt{2}}
\left[ J^{+}_{qL\mu} J^{-\mu}_{qL} + J^{+}_{\ell L\mu}
J^{-\mu}_{\ell L} \right. \nonumber \\  & & +
\frac{1-\eta}{1+\eta} \;  \left. (J^{+}_{qL\mu} J^{-\mu}_{\ell L}
+ J^{-}_{qL\mu} J^{+\mu}_{\ell L}) \right]  .
\end{eqnarray}
Thus, $\eta$ makes the strength of the leptonic charged-current
weak interactions different from the strength of the semileptonic
weak interactions, so that semileptonic processes can be
considered to constrain its value.  We will do this by adapting a
calculation by Rajpoot~\cite{Georgi} based on preserving Cabbibo
universality and Cabbibo-Kobayashi-Maskawa (CKM) unitarity.  The
relevant four fermion effective Lagrangian is given by
\begin{eqnarray}
& \displaystyle {\cal L}^{\mbox{eff}}_{\mbox{CC}}(\nu q) =
\frac{G_{F}}{\sqrt{2}} \; \frac{1-\eta}{1+\eta} \; V_{qq'}
\overline{q'} \gamma_{\mu} (1+\gamma_{5}) q \; \overline{\ell}
\gamma^{\mu} (1+\gamma_{5}) \nu , &
\nonumber \\
& & \label{nothing}
\end{eqnarray}
where $V_{qq'}$ are the elements of the CKM mixing matrix.
Typically these are determined from leptonic decays of hadrons.
The data from high precision measurements yields~\cite{PDG}
\begin{eqnarray}
\frac{1-\eta}{1+\eta} \; |V_{ud}| & = & 0.9738 \pm 0.0005 ,
\nonumber \\
\frac{1-\eta}{1+\eta} \; |V_{us}|  & = & 0.2200 \pm 0.0026 ,
\nonumber \\
\frac{1-\eta}{1+\eta} \; |V_{ub}|  & = & 0.00367 \pm 0.00047 .
\label{unitarity}
\end{eqnarray}
Unitarity of the mixing matrix requires $|V_{ud}|^{2} +
|V_{us}|^{2} + |V_{ub}|^{2} = 1$ .  This constraint immediately
implies $\eta=0.00083 \pm 0.00037$ and $m_{W_{2}}=2.8 \pm 0.9$
TeV, although the experimental results quoted in
Eq.~(\ref{unitarity}) are still controversial \cite{CMS}. It is
worth emphasizing that there is no analogous result in ununified
models because the low-energy charged-current structure differs
from that of the unified variant.

Additional limits can be obtained from constraints at higher
energies coming from precision electroweak measurements.  With the
increasing accuracy of data, a very convenient choice of free
parameters to use in the gauge sector of the extended electroweak
model is given by $\alpha_{em}$, $G_{F}$, $m_{Z_{1}}$ and
$m_{W_{1}}$, since $m_{Z_{1}}$ and $m_{W_{1}}$ are directly
measurable.  Then the other parameters of the theory like
$\sin^{2} \theta_{W}$ and $\eta$ become dependent parameters
defined through Eqs.~(\ref{gs}), (\ref{mass}), and~(\ref{Fermi}),
by
\begin{eqnarray}
& & \sin^{2} \theta_{W} = 1 -  \frac{m_{W_{1}}^{2}}{m_{Z_{1}}^{2}} ,
\nonumber \\
& & 1 + \eta = \frac{\sqrt{2} G_{F}}{\pi \alpha_{em}} m^{2}_{W_{1}}
\sin^{2}\theta_{W} (1 - \Delta r) ,
\label{eta1}
\end{eqnarray}
where the weak radiative correction parameter $\Delta r$ has been
included. Considering that the SM value of $\Delta r$ does not
change at the energy scale of the model which produces $\eta > 0$,
$\eta$ and therefore $m_{Z_{2}}$ (= $m_{W_{2}}$) can be predicted
from $m_{Z_{1}}$ and $m_{W_{1}}$.  However, they are dominated by
the experimental bar errors.  We only get an upper bound $\eta <
0.0024$ which implies a lower mass bound of $m_{Z_{2}(W_{2})} >
1.6$ TeV for the nonstandard weak bosons, consistent with the
above results.  The magnitude of $\Delta r = 0.03434 \mp 0.0017
\pm 0.00014$ used to obtain these limits was extracted from the
minimal SM in the on-shell scheme~\cite{PDG}. All nonstandard
particles were not included in the radiative corrections predicted
by the SM because we search for small tree-level effects of the
extra gauge bosons.

Another limit can be derived from Eq.~(\ref{eta1}) by setting in
the SM prediction for $m_{W_{1}}$ instead of the $\Delta r$ term.
We approximately find
\begin{equation}
\eta = 2 \; \frac{\Delta m_{W_{1}}}{m_{W_{1}}} \;
(\frac{1}{sin^{2} \theta_{W}} - 2) ,
\label{eta2}
\end{equation}
where $\Delta m_{W_{1}}=m^{SM}_{W_{1}}-m_{W_{1}}$, obtaining
forthwith the bounds $\eta < 0.0013$ and $m_{Z_{2}(W_{2})} > 2.2$
TeV.  Equation~(\ref{eta2}) shows how, in the case of the
$W_{1}$-mass, a small deviation of the experimental result from
the SM prediction defines the $\eta$-parameter of the new physics
implied by the $\tilde{\mbox{P}}$-symmetric model. The model
favors an experimental value of $m_{W_{1}}$ below the SM
prediction. Lower mass bounds coming from other observables can be
computed as well.

Thus, more experimental accuracy is needed to have a signal of the
new physics from precision electroweak measurements at high
energies, where the SM exhibits a very good agreement with data.
We know that the SM cannot be complete, so eventually quantitative
disagreements between data and the SM predictions shall occur. We
expect that all the magnitudes will be accounted for in part by a
simple, consistent value of $\eta$, which in turn will give the
mass scale for the new left-handed weak bosons. In a sense, the
$\eta$ parameter plays a role similar to that of the mixing angle
$\theta_{W}$ in the SM prior to the LEP era; its value obtained
from a fit to neutral-current data made possible to predict the
mass of weak bosons.

In conclusion, we have analyzed the unified model
$\mbox{SU(3)}_{q} \times \mbox{SU(3)}_{\tilde{q}} \times
\mbox{SU(2)}_{q \tilde{\ell}} \times \mbox{SU(2)}_{\tilde{q} \ell}
\times \mbox{U(1)}_{Y} \times \tilde{\mbox{P}}$ in which the ${\bf
Z}_{2}$ exotic symmetry defined by $\tilde{\mbox{P}}$ is
spontaneously broken.  It may be viewed as an important stage in
the process of full understanding which can be tested at
relatively low energies. Our aim in this paper has been limited to
account for the observed universality violation of weak
interactions and consider phenomenological constraints on the
masses and mass-splittings of exotic generations of quarks and
leptons that seem to enforce the SM with two Higgs doublets as
well as further gauge extensions of our model. What is interesting
is that the lighter weak bosons that arise have the same
properties as the standard ones, though all predictions of the SM
(starting at the tree level) are corrected by $\eta$, the new weak
parameter introduced by the symmetric model. The magnitude of
$\eta$ actually measures the no universality of weak interactions.
We have fixed the value of this parameter using
universality/unitarity constraints, which sets a mass of order
$2.8$ TeV for the nonstandard weak bosons, even though the
experimental results used are still controversial. Their existence
mean to touch the idea of symmetric left-handed weak forces with
experiment, stressing the open question which inquires for the new
dynamics that may be behind or associated with exotic symmetry.
This work shows, however, that despite the lack of knowledge of
this dynamics, much information can be obtained from this
symmetry.

As a way forward, the notion that exotic symmetry extends to the
forces of the SM already suggests to include the $\mbox{U(1)}_{Y}$
interactions to have the unified gauge group $\mbox{G}_{q
\tilde{\ell}} \times \mbox{G}_{\tilde{q} \ell} \times
\tilde{\mbox{P}}$ with $\mbox{G} = \mbox{SU(3)} \times
\mbox{SU(2)} \times \mbox{U(1)}$.  Next, it is natural to embed
each $\mbox{G}$ in a GUT to have $(\mbox{GUT})_{q \tilde{\ell}}
\times (\mbox{GUT})_{\tilde{q} \ell} \times \tilde{\mbox{P}}$
which unifies standard and exotic quarks and leptons, and also the
strong and electroweak interactions; the $\tilde{\mbox{P}}$
symmetry should be preserved down to energies where the
spontaneous breaking of the electroweak gauge symmetry takes
place.  The ordinary nucleon does not decay in models of this type
because of the separation of GUT for ordinary quarks and leptons.
Thus, applications of GUT techniques~\cite{Mohapatra} predict
transitions between exotic and ordinary quarks and leptons, giving
rise to the decay of heavy exotic matter into ordinary matter,
with separate conservation of ordinary and exotic color charge.
From a cosmological point of view, this is a mandatory condition
not to conflict with observational evidences on the absence of
exotic baryons.  On the other hand, the well-known cosmological
domain wall problems associated with the spontaneous breakdown of
the discrete symmetry $\tilde{\mbox{P}}$ can be solved with the
mechanism of destroying domain walls by primordial black
holes~\cite{SFS}, without constraints on the parameters in the
Lagrangian of the model. Finally, from the viewpoint of precision
electroweak experiments, current constraints on the masses and
mass-splittings of new fermions also compel, within the context of
our model, the duplication of the SM forces with a
two-Higgs-doublet extension included.
\\[5pt] \indent
This work was partially supported by the Departamento de
Investigaciones Cient\'{\i}ficas y Tecnol\'ogicas, Universidad de
Santiago de Chile.

\end{document}